\documentclass[11pt,a4paper]{article}
\usepackage{jheppub}

\usepackage{graphicx}
\usepackage{dcolumn}
\usepackage{bm}
\usepackage{amsmath}
\usepackage{latexsym}
\newcommand{\be}{\begin{equation}}
\newcommand{\ee}{\end{equation}}
\newcommand{\ea}{\end{eqnarray}}
\newcommand{\ba}{\begin{eqnarray}}
\def\ni{\noindent}

\def\[{\left\lbrack}
\def\]{\right\rbrack}

\def\({\left(}
\def\){\right)}

\begin{document}

\title{\Large{Ho\v{r}ava-Lifshitz cosmological models in noncommutative space-times}}

\author[a,b]{E. M. C. Abreu,}
\author[b]{A. C. R. Mendes,}
\author[b]{G. Oliveira-Neto,}
\author[b]{J. Ananias Neto}
\author[b]{L. G. Rezende Rodrigues}
\author[b]{and M. Silva de Oliveira}

\affiliation[a]{Grupo de F\'isica Te\'orica e  F\' isica-Matem\'atica, Departamento de F\'{\i}sica, \\
Universidade Federal Rural do Rio de Janeiro\\
BR 465 km 07, 23890-971, Serop\'edica, RJ, Brazil}
\affiliation[b]{Departamento de F\'{\i}sica, ICE, Universidade Federal de Juiz de Fora,\\
36036-330, Juiz de Fora, MG, Brazil}
\emailAdd{evertonabreu@ufrrj.br}
\emailAdd{albert@fisica.ufjf.br}
\emailAdd{gilneto@fisica.ufjf.br}
\emailAdd{jorge@fisica.ufjf.br}
\emailAdd{lgrr.fisica.ufjf@gmail.com}
\emailAdd{monalisa-silva@hotmail.com}


\abstract{In this work, we will analyze a noncommutative (NC) version of the Friedmann-Robert-Walker cosmological models within the gravitational Ho\v{r}ava-Lifshitz theory. The matter content of the models is described by a perfect fluid and the constant curvature of the spatial sections may be positive, negative or zero. In order to obtain this theory, we will use the Faddeev-Jackiw symplectic formalism to introduce, naturally, space-time noncommutativity inside the equations that provide the dynamics of the theory. We will investigate, in details, the classical field equations of a particular version of the NC models. The equations will be modified, with respect to the commutative ones, by the introduction of a NC parameter. We will demonstrate that various NC models, with different types of matter and spatial constant curvatures, show several interesting and new results relative to the corresponding commutative ones. We will pay special attention to some cases, where the NC model predicts a scale factor accelerated expansion, which may describe the current state of our Universe.}




\maketitle
\flushbottom

\pagestyle{myheadings}
\markright{Noncommutative space-time Ho\v{r}ava-Lifshitz cosmological models}



\section{Introduction}

Ho\v{r}ava $4D$ description for gravity \cite{horava} was inspired by condensed matter systems of dynamical critical structures.  It has explicit $3D$ spatial general covariance and time reparametrization invariance, as we will see in a moment.  However, it only obtains $4D$ general covariance in an IR large distance limit.   Moreover, as we will explore here, it can also be described in a language analogous to the $4D$ Arnowitt, Deser and Misner canonical general relativistic formalism.  We will be back to this issue soon, after some physical considerations and motivations.

If superstrings theory is the correct one to unify all the interactions in nature, the existence of a noncommutative (NC) space-time must have played the dominant role at very early stages of our Universe.
At that time, all the canonical variables and corresponding momenta describing our Universe should have obeyed a NC algebra. Inspired by these ideas some 
researchers have considered such NC models in quantum cosmology \cite{garcia}-\cite{monalisa}. 
It is also possible that some residual NC contributions may have survived at later stages of our Universe. Based on these ideas, some researchers have proposed 
some NC models in classical cosmology in order to explain 
the present accelerated expansion of our Universe \cite{pedram}-\cite{afonso}.

When it is applied to cosmology, noncommutativity (NCY) has produced very important results. So far, those
results were restricted to cosmological models based on general relativity (GR). In situations of exceedingly
intense gravitational interaction, for very general and reasonable conditions, space-times described by 
GR develop singularities. For those situations, GR cannot describe properly 
the gravitational interaction. 
One proposal in order to eliminate those singularities was to quantize GR. 
Unfortunately, it was shown that GR is not perturbatively renormalizable \cite{weinberg0}. 
After that discovery, many geometrical theories of gravity, different from general relativity and perturbatively renormalizable, 
have been introduced. 
Alas, those theories produce massive ghosts in their physical spectrum and are
not unitary \cite{stelle}. 

An interesting example of such geometrical theories of gravity is the 
Ho\v{r}ava-Lifshitz theory (HL)\cite{horava}.
As we said above, this theory has an explicitly asymmetry between space and time, which manifests through an anisotropic scaling 
between space and time. That property means that the Lorentz symmetry is broken, at least at high energies, where that 
asymmetry between space and time takes place. At low energies the HL theory tends to GR, recovering
the Lorentz symmetry. This anisotropic scaling between space and time is given by,
\begin{equation}
\label{1a}
t \to b^z t \qquad \mbox{and} \qquad \vec{x} \to b \vec{x},
\end{equation}

\ni where $b$ is a scale parameter and $z$ is a dynamical critical exponent. As discussed by Ho\v{r}ava \cite{horava}, a theory
of gravity using those ideas is power-counting renormalizable, in 3+1 dimensions, for $z = 3$.  Besides, GR is recovered when $z \to 1$.
Due to the asymmetry between space and time present in his geometrical theory of gravity, Ho\v{r}ava decided to formulate it
by using the Arnowitt-Deser-Misner (ADM) formalism, which splits the four dimensional space-time in a time line plus three dimensional 
space \cite{misner}.   
In the ADM formalism the four  dimensional metric $g_{\mu\nu}$ ($\mu, \nu = 0,1,2,3$) is decomposed in terms
of the three dimensional metric $h_{i j}$ ($i, j = 1,2,3$), of spatial sections, the shift vector $N_i$ and the lapse function $N$, which is viewed as a gauge field for time reparametrizations.   
In general all those quantities depend both on space and time. In his original work, Ho\v{r}ava considered the simplified assumption that 
$N$ should depend only on time \cite{horava}. This assumption has became known as the {\it projectable condition}. Although many works 
have been written about HL theory using the {\it projectable condition}, some authors have considered the implications of working in the 
{\it non-projectable condition}. In other words, they have considered $N$ as a function of space and time \cite{blas,blas1}.
The gravitational action of the HL theory was proposed such that the kinetic component was constructed separately from the potential one.
The kinetic component was motivated by the one coming from GR, written in terms of the extrinsic curvature tensor. It contains
time derivatives of the spatial metric up to the second order and one free parameter ($\lambda$), which is not present in the general relativity 
kinetic component.   At the limit when $\lambda \to 1$, one recovers GR kinetic component. The potential component must
depend only on the spatial metric and its spatial derivatives. As a geometrical theory of gravity, the potential component of the HL theory
should be composed of scalar contractions of the Riemann tensor and its spatial derivatives. 

In his original paper \cite{horava}, Ho\v{r}ava considered a simplification in order to reduce the number of possible terms contributing to the potential component of his theory. It is called the {\it detailed balance condition}. Although this condition indeed reduces the number of terms contributing to the potential component,
some authors have shown that, without using this condition, it is possible to construct a well defined and phenomenologically interesting theory, 
without many more extra terms \cite{mattvisser,mattvisser1}. Like other geometrical theories of gravity, it was shown that the projectable 
version of the HL theory, with the detailed balance condition, have massive ghosts and instabilities \cite{mattvisser1,wang}. The HL theory have been applied to cosmology and produced very interesting models \cite{wang1,bertolami,paul,kord,pedram2,tawfik,pourhassan,pedram1,giani,gil3}.

In the present paper, we will consider the possibility that some residual NC contribution may have survived all the way until the current stage of the Universe. 
Here, we want to investigate how such residual NC contributions can modify some commutative classical cosmological models, based on the HL gravitational theory. 
We are interested in the modifications that may help to explain the current accelerated expansion of the Universe \cite{riess,perlmutter}. To obtain this theory, we have used the Faddeev-Jackiw symplectic formalism to introduce, naturally, NCY inside the equations that provide the dynamics of the theory. This formalism have already been used to obtain NC FRW cosmological models, based on GR \cite{mateus,gil4}.

We have organized this paper as follows. In section \ref{section2}, we have constructed the
general commutative Hamiltonian for the FRW cosmological models, coupled to a perfect fluid, based on the HL gravitational theory.
In section \ref{section3}, we have built the
NC versions of several FRW cosmological models, based on the HL gravitational theory, using the Faddeev-Jackiw (FJ) symplectic formalism.
These models may have positive, negative or zero constant spatial curvatures and they
are coupled to different types of perfect fluids. We have used the ADM and Schutz variational formalisms in order to
write the Hamiltonians of the models. 
In section \ref{section4}, we have solved the dynamical 
equations for the scale factor and analyzed the evolutions of the universes, for a particular version of the NC models.
Based on our results, we have concluded how the NC parameter modifies the evolution of the corresponding commutative models. In particular, we have explained how it can be used to describe the present accelerated expansion of the Universe. We have 
presented some explicit examples where our conclusions can be clearly verified. Also in this section, we have shown that
the NC term, present in the dynamical equations, may be interpreted as the one coming from a NC perfect fluid. 
We have provided a numerical estimation for the value of the NC parameter $\alpha$ in 
section \ref{section5}. Our final considerations and conclusions are in the last section, i.e., section \ref{section6}.

\section{The commutative Ho\v{r}ava-Lifshitz Friedmann-Robertson-Walker models}
\label{section2}

FRW cosmological models are characterized by the scale factor $a(t)$ and have the following line element,

\begin{equation}
\label{1}
ds^2 = - N(t)^2 dt^2 + a(t)^2\left( \frac{dr^2}{1 - kr^2} + r^2 d\Omega^2
\right)\, ,
\end{equation}

\ni where $d\Omega^2$ is the line element of the two-dimensional sphere with
unitary radius, $N(t)$ is the lapse function \cite{wheeler} and $k$ represents the
constant curvature of the spatial sections. The curvature is positive for $%
k=1$, negative for $k=-1$ and zero for $k=0$. Here, we are using the natural
unit system, where $c = 8\pi G = 1$.  We assume that the matter content of the model is
represented by a perfect fluid with four-velocity $U^\mu = \delta^{\mu}_0$
in the co-moving coordinate system used. The energy-momentum tensor is given by,

\begin{equation}
T_{\mu\nu} = (\rho+p)U_{\mu}U_{\nu} + p g_{\mu\nu} ,
\label{2}
\end{equation}
where $\rho$ and $p$ are the energy density and pressure of the fluid,
respectively. The Greek indices $\mu$ and $\nu$ run from zero to three. The equation of state for a perfect fluid is $p = \omega\rho$.

The action for the projectable HL gravity, without the detailed balance condition, for $z=3$ and in $3+1$-dimensions is given by \cite{bertolami},

\begin{eqnarray} \label{3}
\mathcal{S}_{HL} & = & \frac{M_{p}^{2}}{2} \int d^{3}x dt N \sqrt{h}  \left[ K_{ij}K^{ij} - \lambda K^{2} - g_{0}{M_{p}}^{2} - g_{1} R  - {M_{p}}^{-2}\Big( g_{2}R^{2} + g_{3}R_{ij}R^{ij} \Big) \right. \nonumber \\
& - &  \left. {M_{p}}^{-4} \left( g_{4}R^{3} + g_{5}R R^{\,i}_{\phantom{i}j} R^{\,j}_{\phantom{j}i} + g_{6} R^{\,i}_{\phantom{i}j} R^{\,j}_{\phantom{j}k} R^{\,k}_{\phantom{k}i} + g_{7} R \nabla^{2} R + g_{8} \nabla_{i}R_{jk} \nabla^{i}R^{jk} \right) \right], \nonumber \\
\end{eqnarray}

\ni where $g_i$ and $\lambda$ are parameters associated with HL gravity, $M_p$ is the Planck mass,
$K_{ij}$ are the components of the extrinsic curvature tensor and $K$ represents its trace,
$R_{ij}$ are the components of the Ricci tensor computed for the metric of the spatial
sections $h_{ij}$, $R$ is the Ricci scalar computed for $h_{ij}$, $h$ is the determinant of
$h_{ij}$ and $\nabla_i$ represents covariant derivatives. The Latin indices $i$ and $j$ run from one to three. As we have mentioned above, GR is recovered at the limit $\lambda \to 1$.

Introducing the metric of the spatial sections that comes from the FRW space-time (\ref{1}), in the action (\ref{3}) and using the following choices: $g_{0} M_{p}^{2} = 2 \Lambda$ and $g_{1} = -1$, we can write that,
\begin{eqnarray}
\label{4}
\mathcal{S}_{HL} & = & \kappa \int_{}^{} dt N \left[ - \frac{\dot{a}^{2} a}{N^{2}} + \frac{1}{3 \left(3 \lambda - 1 \right)} \left( 6k a - 2\Lambda a^{3} - \frac{12k^{2}}{a M_{p}^{2}} \left( 3g_{2} + g_{3} \right) \right. \right. \nonumber \\
& - & \left. \left. \frac{24k^{3}}{a^{3} {M_{p}^{4}}} \left(9g_{4} + 3g_{5} + g_{6} \right) \right) \right],
\end{eqnarray}
where $$\kappa = \frac{3(3\lambda-1) M_{p}^{2} }{2} \int_{}^{} d^{3}x \frac{r^{2} sen \theta}{\sqrt{1 - kr^{2}}}\,\,.$$ If we choose, for simplicity, $\kappa=1$, we will write the HL Lagrangian density ($\mathcal{L}_{HL}$), from $\mathcal{S}_{HL}$ in Eq. (\ref{4}) as,
\begin{equation}
\label{5}
\mathcal{L}_{HL} = N \left[ -  \frac{\dot{a}^{2}a}{N^{2}} + g_{c}ka - g_{\Lambda} a^{3} - g_{r}\frac{k^{2}}{a} - g_{s} \frac{k^{3}}{a^{3}} \right],
\end{equation}
where the new parameters are defined by,
\begin{eqnarray}
\label{6}
g_{c} & = & \frac{2}{3 \lambda - 1}, \quad
g_{\Lambda} = \frac{2 \Lambda}{3 \left( 3 \lambda - 1 \right)}, \quad
g_{r} = \frac{4}{(3\lambda-1)M_p^2} \left( 3g_{2} + g_{3} \right), \\\nonumber
g_{s} & = & \frac{8}{(3\lambda-1)M_p^4} \left(9 g_{4} + 3g_{5} + g_{6} \right).
\end{eqnarray}
The parameter $g_c$ is positive, by definition, and the others may be positive or negative.

Now, we want to write the HL Hamiltonian density.   To accomplish the task, we must
compute the momentum canonically conjugated to the single dynamical variable, present in the geometry sector, i.e., the scale factor. Using the definition, that momentum ($P_a$) is
given by,
\begin{equation}
\label{7}
P_{a} = \frac{\partial \mathcal{L}}{\partial \dot{a}} = \frac{\partial}{\partial \dot{a}} \left[ - \frac{\dot{a}^{2}a}{N} \right] = - \frac{2 \dot{a} a}{N}\,\,.
\end{equation}
Introducing $P_a$ (\ref{7}) into the definition of the Hamiltonian density, with the aid of $\mathcal{L}_{HL}$ (\ref{5}), we obtain, the following HL Hamiltonian ($H_{HL}$),
\begin{equation}
H_{HL}=N\mathcal{H}_{HL} = N \left[ - \frac{P_{a}^{2}}{4 a} - g_{c}ka + g_{\Lambda} a^{3} + g_{r}\frac{k^{2}}{a} + g_{s} \frac{k^{3}}{a^{3}} \right]. 
\label{8}
\end{equation}

In this work, we will obtain the perfect fluid Hamiltonian ($H_{pf}$) using
the Schutz's variational formalism \cite{schutz,schutz1}. In this formalism the four-velocity ($U_\nu$) of the fluid is expressed in terms of six thermodynamical potentials ($\mu$, $\epsilon$, $\zeta$, $\beta$, $\theta$, $S$), in the following way,

\be
\label{9}
U_\nu = \frac{1}{\mu}\left(\epsilon_{,\nu}+\zeta\beta_{,\nu}+\theta S_{,\nu}\right)\,\,,
\ee
where $\mu$ is the specific enthalpy, $S$ is the specific entropy,
$\zeta$ and $\beta$ are connected to rotation and they are absent from the FRW models. 
Finally, $\epsilon$ and $\theta$ have no clear physical meaning. The four-velocity obeys the normalization condition,

\be
\label{10}
U^\nu U_\nu = -1.
\ee

The starting point, in order to write the $H_{pf}$ for the perfect fluid, is the action ($\mathcal{S}_{pf}$), which in this formalism is written as,

\be
\label{11}
\mathcal{S}_{pf} = \int d^4x\sqrt{-g}(16\pi p)\,\,,
\ee
where $g$ is the determinant of the four-dimensional metric ($g_{\alpha \beta}$) and $p$ is the fluid pressure. Introducing the metric (\ref{1}), Eqs. (\ref{9}) and (\ref{10}), the state equation of the fluid and the first law of thermodynamics into the action (\ref{11}), and after some thermodynamical considerations, the action takes the form \cite{germano1},

\be
\label{12}
\mathcal{S}_{pf}=\int dt\left[ 
N^{-1/\omega}a^3\frac{\omega(\dot{\epsilon}+\theta\dot{S})^{1+1/\omega}}{(\omega+1)^{1+1/\omega}}e^{-S/\omega}\right].
\ee

\ni From this action, we can obtain the perfect fluid Lagrangian density and write the Hamiltonian ($H_{pf}$),
\be
\label{13}
H_{pf}=N{\mathcal{H}_{pf}}=N\left(P_{\epsilon}^{\omega+1}a^{-3\omega}e^S\right),
\ee
where $$P_\epsilon = N^{-1/\omega}a^3(\dot{\epsilon}+\theta\dot{S})^{(\omega+1)^{-1/\omega}/\omega}e^{-S/\omega}\,\,.$$
We can further simplify the Hamiltonian (\ref{13}), by performing the following canonical transformations \cite{germano1},

\be
\label{14}
T = -P_S e^{-S}P_\epsilon^{-(\omega+1)},\quad P_T = P_\epsilon^{\omega+1}e^S,\quad \bar{\epsilon} = \epsilon-(\omega+1)\frac{P_S}{P_\epsilon},\quad \bar{P_\epsilon} = P_\epsilon,
\ee
where $P_S = \theta P_\epsilon$. With these transformations the Hamiltonian (\ref{13}) takes the form,

\begin{equation}
H_{pf}=N {\mathcal{H}_{pf}}= N\frac{P_T}{a^{3\omega}},  
\label{15}
\end{equation}
where $P_{T}$ is the momentum canonically conjugated to $T$. We can write now
the total Hamiltonian of the model ($H$), which is written as the sum of $H_{HL}$ (\ref{8}) with $H_{pf}$ (\ref{15}),
\begin{equation}
H =N{\mathcal{H}} = N \left[ - \frac{P_{a}^{2}}{4 a} - g_{c}ka + g_{\Lambda} a^{3} + g_{r}\frac{k^{2}}{a} + g_{s} \frac{k^{3}}{a^{3}} + \frac{P_{T}}{a^{3 \omega}} \right]. 
\label{16}
\end{equation}

The classical dynamics is governed by the Hamilton's equations, derived from
eq. (\ref{16}). In particular, the variation of $H$ with respect to $N$ gives rise to the superhamiltonian constraint equation,
\begin{equation}
\label{15.5}
{\mathcal{H}} = 0.
\end{equation}

\ni which is a constraint equation.  In the next section we will analyze the NC version of the FRW HL model. As a Hamiltonian system, we will analyze its constraints from the FJ point of view.

\section{The NC Ho\v{r}ava-Lifshitz Friedmann-Robertson-Walker models and \\ Faddeev-Jackiw formalism}
\label{section3}

It is very important to mention that the NCY that
we are about to propose is not the usual NCY between usual spatial
coordinates.  
We are describing the FRW models using a Hamiltonian formalism, therefore
the phase spaces of the present models are given by both the canonical variables and conjugated
momenta: $\{ a, P_a, T, P_T \}$. Then, at the classical level, the NC algebra we 
are about to propose will involve  these phase space variables. Since these variables 
are functions of the time coordinate $t$, this procedure is a generalization of the standard 
NCY between usual spatial coordinates. The NCY between those
types of phase space variables have already been proposed in the literature. At the quantum
level in Refs. \cite{garcia}-\cite{monalisa} and at the classical level in
Refs. \cite{pedram}-\cite{afonso}.

In this section we will introduce the NC algebra into the commutative model explored in the last section.  The objective is to obtain the equations of motions of the NC cosmological model to analyze its behavior and consequences.

\subsection{The Faddeev-Jackiw formalism}

To begin to explore the Faddeev-Jackiw (FJ) formalism, we have to clarify that, when we have to deal with constrained
systems, we find certain consistency problems that obliterate the Poisson brackets algebra and consequently
any quantization procedure. One of the seminal papers that elucidate this kind of
problem was accomplished by Dirac in \cite{1}, where a consistent quantization method was introduced for any system. Some years after that, FJ \cite{2} studied constrained systems via symplectic approach applied to first-order Lagrangians. 
Afterward, the so-called FJ symplectic approach was used and it was modified accordingly to several and different systems and purposes.

The FJ method is an approach that is geometrically motivated. It is
dependent of the phase-space symplectic structure. 
Its first-order intrinsic structure allows us to construct the Hamilton equations of motion from a variational principle. 
This geometric structure relative to the Hamiltonian phase-space, can be built directly from the equations of motion that follows from the inverse of the so-called symplectic 2-form, if the inverse exists. 

Just few years after its publication, the FJ formalism was modified by Barcelos-Neto and Wotzasek in \cite{3-4}, by Montani and Wotzasek in \cite{6}, and through the years it has been used in different systems \cite{todosnossos,7}.

This underlying FJ's geometric structure can be realized directly from the elements of the inverse
symplectic matrix.  These elements agree with the corresponding Dirac brackets, which provides a well established connection to the commutators of the quantized theory. 
The results obtained by using the FJ formalism have been compared to the corresponding results from the Dirac method in different scenarios, for both unconstrained and constrained systems.  Because of the multiplicity of situations and applications, it is
still an object of profound investigations.

Technically, we can explain that the FJ method has, as a key ingredient, that the constraints of the object system 
produce deformations in the 2-form symplectic matrix.  
It occurs in such a way that, when all the constraints of the theory are considered (through a Darboux transformation), the symplectic matrix is non-singular. Namely, differently from the Dirac method, we do not classify the constraints in first or second class, they are just constraints, resulting in Dirac brackets.
However, sometimes it occurs that the iteratively deformed 2-form matrix is singular and no new constraints can be obtained from the corresponding zero-mode. 
In this case, we can say that we have a gauge theory. 
At this point we must introduce convenient
gauge (subsidiary) conditions, which act as constraints.
Consequently the 2-form matrix becomes invertible. 
It is basically the heart of Dirac's work, where the stability
of the constraints under time evolution is mandatory. 
In other words, constraints are not solved but embedded in an extended phase space. 
The FJ approach can be considered extremely suitable when some relevant symmetries have to be preserved.

\subsection{The cosmological problem}

As we have just said, above, the NC version of our model will be obtained through the FJ formalism that introduces a NC algebra.  Let us begin with the total Hamiltonian (\ref{16}), 
\begin{eqnarray}
H = N \left[- \frac{P^{2}_{a}}{4a} - g_{c}ak + g_{\Lambda}a^{3} + k^{2}\frac{g_{r}}{a} + k^3\frac{g_{s}}{a^{3}} + \frac{P_{T}}{a^{3\omega}}\right]. 
\label{HS}
\end{eqnarray}

From the above total Hamiltonian, let us begin to use the formalism to obtain the Lagrangian density of the system, which will be called, in this iterative procedure, as the zero-Lagrangian,
\begin{eqnarray}
\mathcal{L}^{\left(0\right)} & = & P_{a}\dot{a} + P_{T}\dot{T} - V^{\left(0\right)}\left(a, P_{a}, T, P_{T}\right) \label{L0}.
\end{eqnarray}

In this last Lagrangian, $V^{\left(0\right)}\left(a, P_{a}, T, P_{T}\right)$, the symplectic potential is given by
\be
V^{\left(0\right)}\left(a, P_{a}, T, P_{T}\right) = N \Omega,
\ee
where
\be
\label{omega}
\Omega  =  - \frac{P^{2}_{a}}{4a} - g_{c}ak + g_{\Lambda}a^{3} + k^{2}\frac{g_{r}}{a} + k^3\frac{g_{s}}{a^{3}} + \frac{P_{T}}{a^{3\omega}}.
\ee

Following the FJ method described in \cite{1,2,3-4,6,todosnossos,7}, the symplectic variables are described by,
\begin{eqnarray}
\xi_{i}^{\left(0\right)} & = & \left(a, P_{a}, T, P_{T}, N\right)\,\,,
\end{eqnarray}

\ni and the 1-form elements in Eq. (\ref{L0}) of the zero-iteration are given by
\be
A_{a}^{\left(0\right)}  =  P_{a}, \qquad
A_{P_{a}}^{\left(0\right)}  =  0 , \qquad
A_{T}^{\left(0\right)}  =  P_{T}, \qquad
A_{P_{T}}^{\left(0\right)}  =  0, \qquad 
A_{N}^{\left(0\right)}  =  0.
\ee

Using the following definition,
\begin{eqnarray}
{f}_{\xi^{i}\xi^{j}} & = & \frac{\partial A_{\xi^{j}}}{\partial \xi^{i}} - \frac{\partial A_{\xi^{i}}}{\partial \xi^{j}} \label{fii},
\end{eqnarray}

\ni we can obtain the matrix elements.  The non-zero elements are,
\begin{eqnarray}
f_{aP_{a}} & = & \frac{\partial A_{P_{a}}}{\partial a} - \frac{\partial A_{a}}{\partial P_{a}} = - 1, \\
f_{P_{a}a} & = & - f_{aP_{a}} = 1, \\
f_{TP_{T}} & = & \frac{\partial A_{P_{T}}}{\partial T} - \frac{\partial A_{T}}{\partial P_{T}} = - 1, \\
f_{P_{T}T} & = & - f_{TP_{T}} = 1, \\ \nonumber
\end{eqnarray}

\ni and all the rest are equal to zero.   Hence, the symplectic matrix for the zero-iteration is
\begin{equation}
f^{\left(0\right)} = \left[\begin{array}{ccccc}
0 & -1 & 0 & 0 & 0\\
1 & 0 & 0 & 0 & 0\\
0 & 0 & 0 & -1 & 0\\
0 & 0 & 1 & 0 & 0 \\
0 & 0 & 0 & 0 & 0 \end{array} \right]\,\,, \label{eqMSI0}
\end{equation}

\ni which is singular.  So, we can carry out the operation,
\begin{equation}
f^{\left(0\right)} = \left[\begin{array}{ccccc}
0 & -1 & 0 & 0 & 0\\
1 & 0 & 0 & 0 & 0\\
0 & 0 & 0 & -1 & 0\\
0 & 0 & 1 & 0 & 0 \\
0 & 0 & 0 & 0 & 0 \end{array} \right].\left[\begin{array}{c}
a \\
b \\
c \\
d \\
e \end{array} \right] = \left[\begin{array}{c}
0 \\
0 \\
0\\
0 \\
0 \end{array} \right], 
\end{equation}

\ni to find the zero mode.  We can see clearly that $e$ remains arbitrary.
So, let us choose conveniently $e = 1$. Hence, the symplectic matrix zero-mode of the zero-iteration (\ref{eqMSI0}) is given by,
\begin{eqnarray}
\nu^{(0)} & = & \left(0 \quad 0 \quad 0 \quad 0 \quad 1\right). 
\end{eqnarray}

Through the multiplication of the zero-mode with the symplectic potential gradient, we obtain another constraint, i.e.,
\begin{eqnarray}
\sum_{i = 1}^{5} \nu^{(0)}_{i}\frac{\partial V^{\left(0\right)}}{\partial \xi^{i}} & = & \nu_{N}\frac{\partial (N \Omega)}{\partial N} = \Omega = 0. \label{vinculo1}
\end{eqnarray}

Following the procedure, such constraint must be introduced into the kinetic part of the first order Lagrangian density, namely, the first iteration Lagrangian 
\begin{eqnarray}
\mathcal{L}^{\left( 1\right)} & = & P_{a}\dot{a} + P_{T}\dot{T} + \Omega\dot{\tau} - V^{\left(1\right)}\left(a, P_{a}, T, P_{T}\right) \label{L1}.
\end{eqnarray}

Now we have that, $V^{\left(1\right)}\left(a, P_{a}, T, P_{T}\right) = V^{\left(0\right)}\left(a, P_{a}, T, P_{T}\right) = N \Omega$, $\tau$ is a Lagrange multiplier and
$\xi_{i}^{\left(1\right)} = \left(a, P_{a}, T, P_{T}, N, \tau\right)$ are the first iteration set of symplectic variables of the system.  The 1-form canonical momenta of this first iteration can be identified as,
\begin{eqnarray}
A_{a}^{\left(1\right)} & = & P_{a}, \qquad 
A_{P_{a}}^{\left(1\right)}  =  0,\qquad 
A_{T}^{\left(1\right)}  =  P_{T}, \\
A_{P_{T}}^{\left(1\right)} & = & 0, \qquad 
A_{N}^{\left(1\right)}  =  0, \qquad
A_{\tau}^{\left(1\right)}  =  \Omega \,\,
\end{eqnarray}

\ni and, with these informations together with the definitions given in (\ref{fii}) we 
can obtain the following symplectic matrix of the first iteration
\begin{equation}
f^{\left(1\right)} = \left[\begin{array}{cccccc}
0 & -1 & 0 & 0 & 0 & \frac{\partial \Omega}{\partial a} \\
1 & 0 & 0 & 0 & 0 & \frac{\partial \Omega}{\partial P_{a}} \\
0 & 0 & 0 & -1 & 0 & 0 \\
0 & 0 & 1 & 0 & 0 & \frac{\partial \Omega}{\partial P_{T}} \\
0 & 0 & 0 & 0 & 0 & 0 \\
- \frac{\partial \Omega}{\partial a} & - \frac{\partial \Omega}{\partial P_{a}} & 0 & - \frac{\partial \Omega}{\partial P_{T}} & 0 & 0 \end{array} \right]\,\,,
\end{equation}

\ni which is singular.  The new zero mode can be calculated from
\begin{equation}
f^{\left(1\right)} = \left[\begin{array}{cccccc}
0 & -1 & 0 & 0 & 0 & \frac{\partial \Omega}{\partial a} \\
1 & 0 & 0 & 0 & 0 & \frac{\partial \Omega}{\partial P_{a}} \\
0 & 0 & 0 & -1 & 0 & 0 \\
0 & 0 & 1 & 0 & 0 & \frac{\partial \Omega}{\partial P_{T}} \\
0 & 0 & 0 & 0 & 0 & 0 \\
- \frac{\partial \Omega}{\partial a} & - \frac{\partial \Omega}{\partial P_{a}} & 0 & - \frac{\partial \Omega}{\partial P_{T}} & 0 & 0 \end{array} \right].\left[\begin{array}{c}
a \\
b \\
c \\
d \\
e \\
f \end{array} \right] = \left[\begin{array}{c}
0 \\
0 \\
0 \\
0 \\
0 \\
0 \end{array} \right], 
\end{equation}

We will choose conveniently that $e = 1$ and $f = 1$. Hence, we can obtain the following zero-mode of the first iteration
\begin{eqnarray}
\nu^{\left(1\right)} = \left( - \frac{\partial \Omega}{\partial P_{a}} \quad \frac{\partial \Omega}{\partial a} \quad - \frac{\partial \Omega}{\partial P_{T}} \quad 0 \quad 1 \quad 1\right).  
\end{eqnarray}

\ni The multiplication of the first iteration zero-mode with that potential gradient is given by
\begin{eqnarray}
\sum_{i = 1}^{6} \nu_i^{(1)}\frac{\partial V^{\left(1\right)}}{\partial \xi^{i}} & = & - N\frac{\partial\Omega}{\partial P_{a}}\frac{\partial \Omega}{\partial a} + N\frac{\partial \Omega}{\partial a}\frac{\partial \Omega}{\partial P_{a}} + \Omega  = 0 \nonumber \\
&\Rightarrow& \Omega  =  0.
\end{eqnarray}

\ni As we can see the first-iteration zero-mode ($\nu^{\left(1\right)}$) provides the same constraint previously obtained in
 (\ref{vinculo1}).  This result indicates that the system has a gauge symmetry that must be fixed and introduced into the zero-Lagrangian
 (\ref{L0}),  in agreement with the symplectic method \cite{gil4}. In order to fix this symmetry, we introduce a gauge fixing term
$(\Sigma)$ into the original zero-Lagrangian (\ref{L0}), 
\begin{eqnarray}
\mathcal{L}^{\left(0\right)} & = & P_{a}\dot{a} + P_{T}\dot{T} + \Sigma\dot{\eta} - V^{\left(0\right)}\left(a, P_{a}, T, P_{T}\right) .\label{eqL}
\end{eqnarray}
with
\begin{eqnarray}
\Sigma & = & N - 1,
\end{eqnarray}
\ni which implies that $N = 1$. In the Lagrangian density written just above
 $V^{\left(0\right)}\left(a, P_{a}, T, P_{T}\right) = N \Omega$, $\eta$ is a Lagrange multiplier and $\xi_{i}^{\left(0\right)} = \left(a, P_{a}, T, P_{T}, N, \eta \right)$, which are the symplectic variables of the system.

If we use the results obtained above
we can write the symplectic matrix of the zero iteration as,
\begin{equation}
f^{\left(0\right)} = \left[\begin{array}{ccccccc}
0 & -1 & 0 & 0 & 0 & 0 \\
1 & 0 & 0 & 0 & 0 & 0 \\
0 & 0 & 0 & -1 & 0 & 0 \\
0 & 0 & 1 & 0 & 0 & 0 \\
0 & 0 & 0 & 0 & 0 & 1 \\
0 & 0 & 0 & 0 & -1 & 0 \end{array} \right]\,\,,
\end{equation}

\ni which is non-singular.  Consequently we can compute its inverse as being,
\begin{eqnarray}
\left[f^{\left(0\right)}\right]^{-1} & = & \left[\begin{array}{ccccccc}
0 & 1 & 0 & 0 & 0 & 0 \\
-1 & 0 & 0 & 0 & 0 & 0 \\
0 & 0 & 0 & 1 & 0 & 0 \\
0 & 0 & -1 & 0 & 0 & 0 \\
0 & 0 & 0 & 0 & 0 & -1 \\
0 & 0 & 0 & 0 & 1 & 0 \end{array} \right]. \label{eq: invers}
\end{eqnarray}

\ni which elements are the Poisson brackets of the theory.   This is the standard method.  But we are interested in the introduction of a NC algebra.  We will do that in this last matrix and carry out the procedure in a reverse path to obtain the NC space-time Lagrangian.

So, let us begin by the NC algebra assuming the following relations between the nonzero Poisson brackets,
\begin{eqnarray}
\left\{a, T\right\} & = & \sigma, \label{P1} \\
\left\{P_{a}, P_{T}\right\} & = & \alpha, \label{P2} \\
\left\{a, P_{T}\right\} & = & \gamma, \label{P3} \\
\left\{T, P_{a}\right\} & = & \chi, \label{P4} \\
\left\{a,P_{a}\right\} & = & \left\{T,P_{T}\right\}=1,\label{P5}
\end{eqnarray}

\ni which must be introduced into \eqref{eq: invers}.
In this way, the inverse of the symplectic matrix  
 (\ref{eq: invers}) including the Poisson brackets 
 (\ref{P1})-(\ref{P5}) can be written as,
\begin{eqnarray}
\left[f^{\left(0\right)}\right]^{-1} & = & \left[\begin{array}{ccccccc}
0 & 1 & \sigma & \gamma & 0 & 0 \\
-1 & 0 & -\chi & \alpha & 0 & 0 \\
-\sigma & \chi & 0 & 1 & 0 & 0 \\
-\gamma & -\alpha & -1 & 0 & 0 & 0 \\
0 & 0 & 0 & 0 & 0 & -1 \\
0 & 0 & 0 & 0 & 1 & 0 \end{array} \right]. \label{eq: inc}
\end{eqnarray}

\ni and the symplectic matrix is given by
\begin{eqnarray}
f^{(0)} = \frac{1}{\Gamma}\left[\begin{array}{ccccccc}
0 & 1 & -\alpha & -\chi & 0 & 0 \\
-1 & 0 & \gamma & -\sigma & 0 & 0 \\
\alpha & -\gamma & 0 & 1 & 0 & 0 \\
\chi & \sigma & -1 & 0 & 0 & 0 \\
0 & 0 & 0 & 0 & 0 & \Gamma \\
0 & 0 & 0 & 0 & - \Gamma & 0 \end{array} \right] \label{eqMSNC}. 
\end{eqnarray}
\ni where $\Gamma = \left(\alpha \sigma - 1\right) + \chi \gamma$ and $\left(\alpha \sigma - 1\right) + \chi \gamma \neq 0$.

In order to proceed with the method, we will use the symplectic matrix elements (\ref{eqMSNC}) and the relations in Eq. (\ref{fii}).  The result is a system of partial differential equations,
\begin{eqnarray}
f_{P_a a} & = & f_{P_T T} \,=\,\frac{1}{\Gamma}, \label{ta1} \\
f_{T a} & = & - \frac{\alpha}{\Gamma}, \label{ta2}\\
f_{P_T a} & = & - \frac{\chi}{\Gamma}, \label{ta3} \\
f_{P_T P_a} & = & - \frac{\sigma}{\Gamma}, \label{ta4} \\
f_{T P_a} & = & \frac{\gamma}{\Gamma}, \label{ta5} \\
f_{\eta N} & = & 1. \label{ta7}
\end{eqnarray}

\ni We will solve the system of partial differential equations (\ref{ta1}-\ref{ta7}), taking in account the following, more general, Lagrangian density,
\begin{eqnarray}
\mathcal{L}_{NC} & = & A_{a}\dot{a} + A_{P_{a}}\dot{P_{a}} + A_{T}\dot{T} + A_{P_{T}}\dot{P_{T}} + A_{\eta}\dot{\eta} + A_{N}\dot{N} - N\Omega. \label{eqLG}
\end{eqnarray}

\ni Therefore, after some algebra to solve the system  (\ref{ta1}-\ref{ta7}), we have that
\begin{eqnarray}
\mathcal{L}_{NC} & = & \frac{1}{\Gamma}\left(P_{a} + 2\alpha T + 2\chi P_{T}\right)\dot{a} + \frac{1}{\Gamma}\left(2a + \gamma T \right)\dot{P_{a}} \nonumber \\ 
& + & \frac{1}{\Gamma}\left(\alpha a + 2\gamma P_{a} + P_{T}\right)\dot{T} + \frac{1}{\Gamma}\left(\chi a + 2 T\right)\dot{P_{T}} \nonumber \\
& + & \eta \dot{N} + \left(2N + p\right)\dot{\eta} - N\Omega. \label{eq: LNC1}
\end{eqnarray}

\ni The second order velocity terms can be eliminated after an integration by parts, hence,
\be
\label{LNC}
\mathcal{L}_{NC}  =  \tilde{P_{a}}\dot{a} + \tilde{P_{T}}\dot{T} + \Sigma\dot{\eta} - N\Omega,
\ee

\ni where we will define the following coordinate transformations in the classical phase space,
\begin{eqnarray}
\label{aa.aa.aa}
\tilde{a} & = & a,  \label{eq: aNC}\\
\tilde{T} & = & T,  \label{eq: TNC}\\
\tilde{P_{a}} & = & \frac{1}{\Gamma}\left( - P_{a} + 2\alpha T + \chi P_{T} \right) , \label{eq: PaNC} \\
\tilde{P_{T}} & = & \frac{1}{\Gamma}\left(\alpha a + \gamma P_{a} - P_{T} \right). \label{eq: PTNC}
\end{eqnarray}
The motivation to introduce the new set of variables $\{\tilde{a}, \tilde{T}, \tilde{P_a}, \tilde{P_T}\}$, is that they satisfy, up to first order in the NC parameters 
($\sigma$, $\alpha$, $\gamma$ and $\chi$) the usual Poisson brackets algebra. When the theory is written in term of those variables, the NCY is described, 
entirely, by the NC parameters. It is important to notice that the transformations leading to the new variables (\ref{eq: aNC}-\ref{eq: PTNC}), were naturally
 derived from the Lagrangian (\ref{LNC}). It means that the FJ formalism induces, naturally, a set of {\it commuting} variables.
%
%
%
%
%
%
\ni From the transformations, induced by the FJ formalism, (\ref{eq: aNC}) and  (\ref{eq: TNC}), it is easy to see that,
$\left\{ \tilde{a}, \tilde{T} \right\}  =  \left\{ a, T \right\} = \sigma$.
Therefore, the new variables $\tilde{a}$ and $\tilde{T}$ will satisfy the usual Poisson brackets algebra only if $\sigma = 0$. Taking that result in account,
we will set $\sigma = 0$, which means that $\Gamma = \chi \gamma - 1$.

\ni Finally, the modified superhamiltonian of the system is identified as being the symplectic potential $N\Omega$, where $\Omega$ is given by Eq. (\ref{omega}). We may
write that superhamiltonian in terms of the new variables if we invert equations (\ref{eq: aNC}-\ref{eq: PTNC}). Therefore, in the present gauge $N=1$, it leads to,
\begin{eqnarray}
\tilde{\mathcal{H}} & = & - \frac{1}{4a}\left(\tilde{P_{a}} + \chi \tilde{P_{T}} + 2 \alpha T\right)^{2} - kg_{c}a + g_{\Lambda}a^{3} + k^{2}\frac{g_{r}}{a} + k^{3}\frac{g_{s}}{a^{3}} \nonumber \\
& + & \frac{1}{a^{3\alpha}}\left(\tilde{P_{T}} + \gamma \tilde{P_{a}} + \alpha a\right). \label{eqHNCa}
\end{eqnarray}
Notice that when $\chi=\alpha=\gamma=0$ in Eqs. (\ref{P2}-\ref{P4}), (\ref{eq: PaNC}) and (\ref{eq: PTNC}), we recover, from (\ref{eqHNCa}), the superhamiltonian $\mathcal{H}$ in (\ref{16}).

\section{Classical behavior of the NC cosmological models}
\label{section4}

\subsection{Dynamical Equations}
\label{subsection4.1}

To investigate the contributions coming from the NC algebra involving the canonical variables and momenta in the classical FRW HL cosmological models, 
we derive the dynamical equations after the computation of the Hamilton's equations from the
total Hamiltonian $N \tilde{\mathcal{H}}$, where $N$ is the lapse function and $\tilde{\mathcal{H}}$ is the modified superhamiltonian Eq. (\ref{eqHNCa}). We also use the constraint equation $\tilde{\mathcal{H}} = 0$. The new momenta $\widetilde{P}_a$ and $\widetilde{P}_{T}$, present in $\tilde{\mathcal{H}}$, are 
given by Eqs. (\ref{eq: PaNC}) and (\ref{eq: PTNC}). In the expressions for $\widetilde{P}_a$ and $\widetilde{P}_{T}$,
$\chi$, $\alpha$ and $\gamma$ are the parameters associated with the NC algebra involving both the canonical variables and momenta in Eqs. (\ref{P2}-\ref{P4}). 
The general case, with all the NC parameters different from zero, is very complicated and we will not treat it here. We hope to return to that case in a future work. In the present analysis, we are going to consider only the contribution coming from the parameter $\alpha$. In other words, we will fix $\chi=\gamma=0$. Another motivation to make this simplification is the possibility of comparing our results with the results of Ref. \cite{mateus}. There, some of us used the same NC algebra in FRW GR cosmological models, coupled to both a perfect fluid and a cosmological constant.

The Hamilton's equations of motion are obtained using the total Hamiltonian $\tilde{H}=N\tilde{\mathcal{H}}$, where $\tilde{\mathcal{H}}$ is given in Eq. (\ref{eqHNCa}), after we set $\chi=\gamma=0$.
They are given, up to fist order in the NC parameter $\alpha$, by
\begin{eqnarray}
\dot{a} & = & - \frac{1}{2a}\left(\tilde{P_{a}} + 2\alpha T\right), \label{GLT1} \\
\dot{\tilde{P_{a}}} & = & - \frac{1}{4a^{2}}\left(\tilde{P_{a}}^{2} + 4\alpha\tilde{P_{a}} T \right) + kg_{c} - 3g_{\Lambda}a^{2} + k^{2}\frac{g_{r}}{a^{2}} + 3k^{3}\frac{g_{s}}{a^{4}} \nonumber \\
& + & \frac{3\omega}{a^{\left(3\omega + 1\right)}}\tilde{P_{T}} + \left(3\omega - 1\right)\frac{\alpha}{a^{3\omega}}, \label{GLT2}\\
\dot{T} & = & \frac{1}{a^{3\omega}}, \label{GLT3} \\
\dot{\tilde{P_{T}}} & = & \frac{\alpha\tilde{P_{a}}}{a} \label{GLT4}\,\,,
\end{eqnarray}
where the dot means time derivative, in the present gauge $N=1$.

Now, we would like to find the dynamical equations concerning the scale factor. This is done in the following way. Using Eqs. (\ref{GLT1})-(\ref{GLT4}), we can obtain the following
equation for $\tilde{P}_T$,
\be
\label{17}
\tilde{P}_T = C - 2\alpha a,
\ee
where $C$ is an integration constant. Physically, for the commutative case ($\alpha=0$), $C$ represents the fluid energy, which means 
that it is positive. Then, using Eq. (\ref{GLT1}), we find, the following equation
expressing $\tilde{P}_a$ in terms of the time derivative of $a$ and $T$,
\be
\label{18}
\tilde{P}_a = -2a\dot{a} - 2\alpha T.
\ee
Finally, we can introduce the values of $\dot{\tilde{P}}_a$ Eq. (\ref{GLT2}), $\dot{T}$ Eq. (\ref{GLT3}), $\dot{\tilde{P}}_T$ Eq. (\ref{GLT4}), 
$\tilde{P}_T$ Eq. (\ref{17}) and $\tilde{P}_a$ Eq. (\ref{18}), in the time derivative of Eq. (\ref{GLT1}). It results, to first
order in $\alpha$, in the following second order, ordinary differential equation for $a$,
\begin{equation}
\label{19}
\ddot{a} + \frac{\dot{a}^{2}}{2a} = - \frac{1}{2a} \left\{ kg_{c} - 3g_{\Lambda}a^{2} + k^{2}\frac{g_{r}}{a^{2}} + 3k^{3}\frac{g_{s}}{a^{4}} +  \frac{3\omega C}{a^{\left(3\omega + 1\right)}} + \frac{\alpha}{a^{3\omega}} \left[ - 3\omega + 1 \right] \right\}.
\end{equation}
As we have mentioned above, there is another dynamical equation for the scale factor which is obtained by imposing the superhamiltonian constraint: $\tilde{\mathcal{H}}=0$, where $\tilde{\mathcal{H}}$ is given in Eq. (\ref{eqHNCa}). That equation is derived by introducing $\tilde{P}_T$ Eq. (\ref{17}) and $\tilde{P}_a$ Eq. (\ref{18}) in the superhamiltonian constraint, 
\begin{eqnarray}
\frac{\dot{a}^{2}}{a^2} + k\frac{g_{c}}{a^2} - g_{\Lambda} - k^{2}\frac{g_{r}}{a^{4}} - k^{3}\frac{g_{s}}{a^{6}} - \frac{C}{a^{3\omega + 3}} + \frac{\alpha}{a^{3\omega + 2}} = 0. \label{20}
\end{eqnarray}
This equation is the generalization, for the NC HL models, of the
Friedmann equation. Both equations (\ref{19}) and (\ref{20}) recover the corresponding dynamical equations of the commutative models when we set $\alpha = 0$.

It will be very useful to rewrite the generalized Friedmann equation (\ref{20})
in the following form,

\begin{equation}
\label{21}
{\dot{a}^2}+V(a)=0,
\end{equation}
where
\begin{equation}
\label{22}
V(a) = kg_{c} - g_{\Lambda}a^{2} - k^{2}\frac{g_{r}}{a^{2}} - k^{3}\frac{g_{s}}{a^{4}} - \frac{C}{a^{3\omega + 1}} + \frac{\alpha}{a^{3\omega}}.
\end{equation}
We notice that the total energy of this conservative system is equal to zero. From the observation of the 
potential curve $V(a)$, we will be able to obtain the qualitative dynamical behavior of $a(t)$.

One important property of the NC HL cosmological models, described by the two dynamical equations above, is that they satisfy the energy conservation equation. 
To show this result, we will use both Eqs. (\ref{19}) and (\ref{21}). We will start by calculating the time derivative of Eq. (\ref{21}),
\begin{equation}
- \left(3\omega + 1\right)\frac{C}{a^{3\omega + 2}}\dot{a} + 3\omega\frac{\alpha}{a^{3\omega + 1}}\dot{a} = 2\frac{\dot{a}}{a}\left(\ddot{a}a - g_{\Lambda}a^{2} + k^{2}\frac{g_{r}}{a^{2}} + 2k^{3}\frac{g_{s}}{a^{4}}\right). 
\label{23}
\end{equation}
If we take the value of $\ddot{a}$ from Eq. (\ref{19}) and introduce it into Eq. (\ref{23}), we obtain that
\begin{eqnarray}
- \left(3\omega + 1\right)\frac{C}{a^{3\omega + 2}}\dot{a} + 3\omega\frac{\alpha}{a^{3\omega + 1}}\dot{a} & = & \frac{\dot{a}}{a}\left(\left\{- {\dot{a}}^{2} - kg_{c} + g_{\Lambda}a^{2} + k^{2}\frac{g_{r}}{a^{2}} + k^{3}\frac{g_{s}}{a^{4}} \right. \right. \nonumber \\
& - & \left. \left. 3\omega \frac{C}{a^{3\omega + 1}} - \left(1 - 3\omega\right)\frac{\alpha}{a^{3\omega}}\right\} \right). 
\label{24}
\end{eqnarray}
Using Eq. (\ref{21}) into the RHS of Eq. (\ref{24}), we can write,
\begin{equation}
\label{25}
- \left(3\omega + 1\right)\frac{C}{a^{3\omega + 2}}\dot{a} = \frac{\dot{a}}{a}\left( - \frac{C}{a^{3\omega + 1}} - 3\omega \frac{C}{a^{3\omega + 1}} \right).
\end{equation}
From the above equation we can see that there is no more contributions coming from the NC parameter $\alpha$. Let us make the assumption that, for a perfect fluid satisfying the equation of state $p=\omega\rho$, the following expression is still true,
\be
\rho = \frac{C}{a^{3\omega + 3}}, 
\label{26}
\ee
to first order in $\alpha$. Therefore, we may write Eq. (\ref{25}) in the following way,
\begin{equation}
\label{27}
\dot{\rho}+3\frac{\dot{a}}{a}(\rho+p)=0.
\end{equation}
This is the energy conservation equation for the commutative version of the models. Therefore, the NCY does not violate the energy conservation law, as expected.






\subsection{Interpreting the noncommutativity as a perfect fluid}
\label{subsection4.2}

In this section, we are going to show how to interpret the contribution originated from
the NCY as a perfect fluid. Before that, we can notice from the generalized Friedmann equation (\ref{20}) that the perfect fluid energy density is given by the term $C/a^{(3\omega+3)}$. Therefore, by comparison, we will propose that the term due to NCY, in Eq. (\ref{20}), must represent a NC perfect fluid ``energy density", with the expression: $-\alpha/a^{(3\omega+2)}$. If we accept this assumption, it would be interesting to know what is the NC perfect fluid equation of state. The best way to do that is by using the energy conservation equation (\ref{27}). Let us call the NC perfect fluid ``energy density" $\vartheta$,
\be
\label{28}
\vartheta = \frac{-\alpha}{a^{(3\omega+2)}}.
\ee
Then, we introduce this quantity $\vartheta$ Eq. (\ref{28}) into Eq. (\ref{27}), which results,
\be
\label{29}
\frac{d}{dt}\left(-\frac{\alpha}{a^{3\omega+2}}\right) + \frac{3}{a}\left(-\frac{\alpha}{a^{3\omega+2}} + p_{NC}\right)\frac{d{a}}{dt} = 0,
\ee
where $p_{NC}$ is the NC perfect fluid ``pressure". After computing the time derivative of the first term in the LHS of Eq. (\ref{29}) and
combining together the resulting terms above, we obtain,
\be
\label{30}
\frac{\dot{a}}{a}\left(\frac{\alpha}{a^{3\omega+2}} \left(3\omega - 1\right) + 3p_{NC}\right) = 0.
\ee
For $\dot{a} \neq 0$ and $a \neq 0$, we can write,
\be
\label{31}
p_{NC} = \left(\omega - \frac{1}{3}\right)\vartheta,
\ee
which is the NC perfect fluid equation of state. We notice that the equation of state (\ref{31}) depends not only on $\alpha$, the NC parameter, but also on $\omega$, which gives the type of perfect fluid coupled to the HL theory.

Observing Eq. (\ref{28}), we see that the ``energy density" of the NC perfect fluid will be positive for $\alpha < 0$, negative for $\alpha > 0$ and zero for $\alpha=0$. For $\alpha < 0$, from Eq. (\ref{31}), we can see that the ``pressure," of this fluid will be positive if $\omega > 1/3$ and negative if $\omega < 1/3$. On the other hand. For $\alpha > 0$, from Eq. (\ref{31}), we see that the ``pressure," of this fluid will be positive if $\omega < 1/3$ and negative if $\omega > 1/3$. Finally, for $\vartheta=0$ or $\omega=1/3$, the ``pressure" of the NC perfect fluid vanishes. In the next section, we will see the general behavior of the solutions, same examples of different types of solutions and their connections to the properties of the NC perfect fluid.


\subsection{Analysis of the models}
\label{subsection4.3}

Now, we want to compute the scale factor dynamics.   To do that, initially, we can observe the potential curve $V(a)$ from Eq. (\ref{22}) and obtain the qualitative behavior of $a(t)$. Then, we will solve Eq. (\ref{19}). Unfortunately, for general values of the
different parameters present in Eq. (\ref{19}), this equation does not have algebraic solutions. Therefore, we will solve it numerically, for each different values of the parameters. 

There are eight parameters in Eq. (\ref{19}). The first one is $k$, which is associated with
the curvature of the spatial sections and it can have three different values: -1, 0, +1. There are other four parameters: $g_c$, $g_\Lambda$, $g_r$ and $g_s$, which are related to the HL theory. With the exception of $g_c$, which is positive, the other three ones are positive or negative. The next one is $\alpha$, the NC parameter, which can be positive, negative or zero. It is important to mention that the last case $\alpha = 0$ means that
the model is commutative. Another parameter is $C$, which is related to the fluid energy density and it is positive. Finally, we have the parameter $\omega$, which is present in the equation of state for the perfect fluid ($p = \omega \rho$). Each value of $\omega$
defines a different perfect fluid. Here, we will consider six different values of $\alpha =
(1, 1/3, 0, -1/3, -2/3, -1)$, which represents respectively: stiff matter, radiation, dust, cosmic strings, domain walls and a cosmological constant. We have solved Eq. (\ref{19}) for all possible values of the parameters.

After solving Eq. (\ref{19}) for all possible cases, mentioned above, we have reached the following general conclusions. If the parameter $\alpha$ is positive, it has the net effect of slowing down the rate of expansion, compared to the commutative case, or even stops it and forces the Universe to contract. If the parameter $\alpha$ is negative, it has the net effect of increasing the rate of expansion, compared to the commutative case. In cases where the Universe is contracting, the presence of a negative $\alpha$ may force the Universe to expand. Therefore, if one wants to describe the present accelerated expansion of our Universe, the models with a negative $\alpha$ are more suitable. Another interesting motivation to study those models with negative $\alpha$, as discussed in Subsection \ref{subsection4.2}, is the fact that for those models the ``energy density" of the NC perfect fluid will be positive. Next, we will present some particular cases where the above conclusions can be clearly verified.

\subsubsection{Expanding universes with $\alpha$ positive, negative and nil}
\label{subsubsection4.3.1}

One of our main motivations, in this work, is to determine whether the NCY, described
here, may explain the current accelerated expansion of our Universe. As explained above, depending
on the sign of the NC parameter $\alpha$, the NCY may speedup ($\alpha < 0$)
or slow down ($\alpha > 0$) the accelerated expansion, compared to the commutative case. Let us
give an example, of this general result, for a specific model where the geometry is
coupled to a perfect fluid of radiation ($\omega=1/3$) and the constant curvature of the
spatial sections is negative ($k=-1$). For this case, the potential $V(a)$ in Eq. (\ref{22}) is
given by,
\begin{eqnarray}
V_{r}(a) = - g_{c} - g_{\Lambda}a^{2} - \frac{\Omega}{a^{2}} + \frac{g_{s}}{a^{4}} + \frac{\alpha}{a}, 
\label{32}
\end{eqnarray}
where $\Omega = g_{r} + C$ and the potential for the commutative case can be obtained by choosing $\alpha=0$ in $V_r$ above. For positive $g_{\Lambda}$ and $\Omega$ and negative
$g_{s}$, from the potential expression, one can see that this Universe starts to
expand from a singularity at $a = 0$. After that, for all values of $\alpha \leq 0$ and the values of $\alpha > 0$, such that the potential $V_r$ (\ref{32}) is negative, 
the scale factor expands initially in a decelerated rate and later on, in an accelerated rate toward $a \to \infty$. If $\alpha$ is positive and the potential $V_r$ (\ref{32}) is negative, the scale factor expands in a rate smaller than in the commutative case. On the
other hand, if $\alpha$ is negative, the scale factor expands in a rate greater
than in the commutative case. An example, of this situation, is shown in Figure 1.

\begin{figure}
\includegraphics[width=8cm,height=6cm, angle=0]{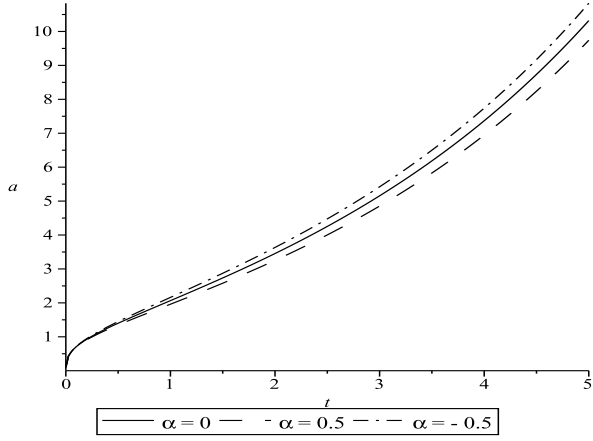}
\caption{Comparison between the scale factor dynamics of three models where the matter is described by radiation and $k = -1$. The first is the commutative model with $\alpha = 0$, the other two are NC models with $\alpha = 0.5$ and
$\alpha = - 0.5$. For all models the parameters have the following values: $g_{c} = 1$, $g_{\Lambda} = 0.1$, $\Omega = 1$ and $g_{s} = -1$.}
\label{fig1}
\end{figure}

\subsubsection{Expanding Universe with $\alpha=0$ versus contracting Universes with $\alpha > 0$}

As we have mentioned above, NCY can modify, in a very fundamental way, the nature of the scale
factor behavior of a given commutative model. Here, we will show one example of this important
result. Let us consider a commutative FRW HL cosmological model where the scale factor
increases as a function of time. After the introduction of the NC algebra, as we will see, the scale
factor will increase up to a maximum value and then it will contract to a {\it big crunch}
singularity. For a specific example of that situation, let us consider a model where the
geometry is coupled to a perfect fluid of stiff matter ($\omega=1$) and the constant
curvature of the spatial sections is negative ($k=-1$). The potential $V(a)$ in Eq. (\ref{22}) is given by,
\begin{equation}
\label{33}
V_{sm}(a) = -g_{c} - g_{\Lambda}a^{2} - \frac{g_{r}}{a^{2}} - \frac{\Omega}{a^{4}} + \frac{\alpha}{a^{3}},
\end{equation}
where $\Omega = - g_s + C$ and the potential for the commutative case can be obtained by choosing $\alpha=0$ in $V_{sm}$ above. For positive $g_{\Lambda}$, $g_{r}$ and $\Omega$, from the potential expression, one may see that this Universe starts to
expand from a singularity at $a = 0$. For the commutative model, the scale factor expands
initially in a decelerated rate and later on in an accelerated rate toward
$a \to \infty$. For the NC space-time model with $\alpha > 0$, let us consider all values of $\alpha$,
such that, the potential $V_{sm}$ (\ref{33}) has a positive maximum. In that case, the scale factor expands up to a maximum size.   Then it is forced to contract toward a {\it big crunch} singularity at $a = 0$. Therefore, we can see that the NC parameter can describe a huge changing in the
evolution of the Universe. An example, of this situation, is shown in Figure 2.

\begin{figure}
\includegraphics[width=8cm,height=6cm, angle=0]{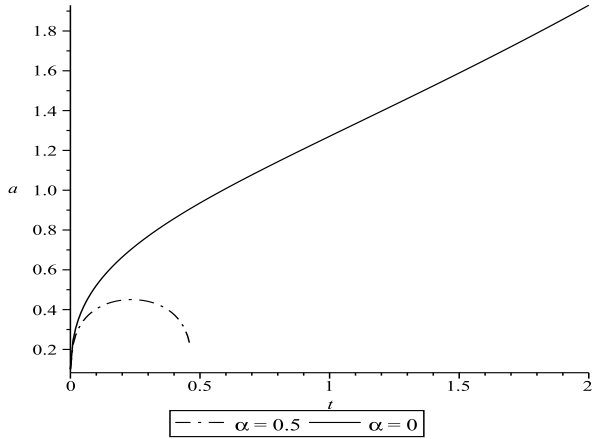}
\caption{Comparison between the scale factor dynamics of two models where the matter is described by stiff matter and $k = -1$. The first is the commutative model with $\alpha = 0$, the other is a NC model with $\alpha = 0.5$. For all models the parameters have the
following values: $g_{c} = 0.1$, $g_{\Lambda} = 0.1$, $g_r = 0.1$ and $\Omega = 0.2$.}
\label{fig2}
\end{figure}

\subsubsection{Expanding Universe with $\alpha < 0$ versus contracting Universes with $\alpha = 0$}

In this case, let us consider a commutative FRW HL cosmological model where the scale factor increases up to a maximum value and then it contracts to a {\it big crunch} singularity. After the introduction of the NC algebra, as we will see, the scale factor will increase
as a function of time. This case is very important because it gives an example of a model where the NCY may describe the accelerated expansion of the Universe. For a
specific example of this situation, let us consider a model where the geometry is coupled
to a perfect fluid of dust ($\omega=0$) and the constant curvature of the spatial sections
is positive ($k=1$). The potential $V(a)$ in Eq. (\ref{22}) is given by,
\begin{equation}
\label{34}
V_d(a) = \Omega - g_{\Lambda}a^{2} - \frac{g_{r}}{a^{2}} - \frac{g_{s}}{a^{4}} - \frac{C}{a}, 
\end{equation}
where $\Omega = g_c + \alpha$ and the potential for the commutative case can be obtained by choosing $\alpha=0$ in $V_d$ above. For positive $g_{\Lambda}$, $g_{r}$, $g_{s}$ and $\Omega$, from the potential expression, one can see that this universe starts to
expand from a singularity at $a = 0$. For the commutative model, if we choose the parameter $g_c$, such that the maximum of the potential $V_d$ in Eq. (\ref{34}) is positive, the scale factor expands up to a maximum size.   Then it is forced to contract toward a {\it big
crunch} singularity at $a = 0$. For the NC model, we will restrict our attention to negative values of $\alpha$
($\alpha < 0$), such that the resulting value of $\Omega = g_c + \alpha$ produces potentials $V_d$ with a negative maximum. In those cases, the scale factor expands initially in a
decelerated rate and later on in an accelerated one toward $a \to \infty$. Therefore, we can see that the NC parameter can describe the evolution of the Universe. In this case, it provides a description of the accelerated expansion of the Universe. An example of this situation is shown in Figure 3.

\begin{figure}
\includegraphics[width=8cm,height=6cm, angle=0]{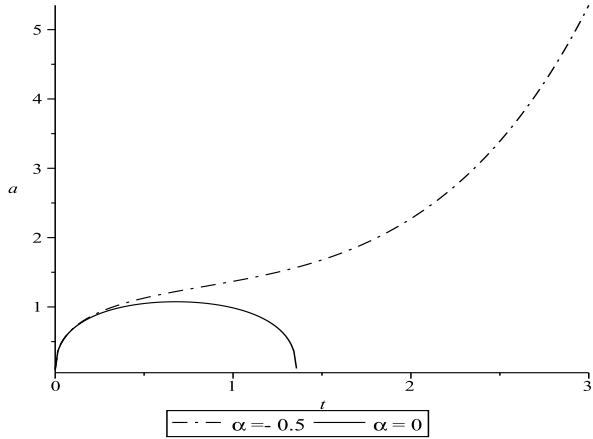}
\caption{Comparison between the scale factor dynamics of two models where the matter is described by dust and $k = 1$. The first one is the commutative model with $\alpha = 0$, the other one is a NC model with $\alpha = -0.5$. For all models the parameters have the
following values: $g_{c} = 3.7$, $g_{\Lambda} = 1$, $g_r = 1$, $g_s = 1$, $C = 1$ and $\Omega = 3.2$.}
\label{fig3}
\end{figure}

\subsubsection{Commutative models without solutions versus NC models with oscillating Universes}

In the current case, let us consider a commutative FRW HL cosmological model where the dynamical equations for the scale factor have no real solutions. After the introduction of the NC algebra, as we will see, the scale factor will increase up to a maximum value and then it will contract to a positive minimum value. After that, it will continue to oscillate between these maxima and minima values. This is an important case because it gives another example of how NCY can modify, in a very fundamental way, the nature of the scale factor behavior of a given commutative model. For a specific example of that situation, let us consider a model where the geometry is coupled to a perfect fluid of cosmic strings ($\omega=-1/3$) and the constant curvature of the spatial sections is negative ($k=-1$). The potential $V(a)$ in Eq. (\ref{22}) is given by,
\begin{equation}
\label{35}
V_{cs}(a) = \Omega - g_{\Lambda}a^{2} - \frac{g_{r}}{a^{2}} + \frac{g_{s}}{a^{4}} + \alpha a\,\,,
\end{equation}
where $\Omega = - g_c - C$ and the potential for the commutative case can be obtained by choosing $\alpha=0$ in $V_{cs}$ above. For positive $g_s$ and negative $\Omega$, $g_{\Lambda}$, $g_{r}$ and $\alpha$, the potential $V_{cs}$ in Eq. (\ref{35}) diverges to $+\infty$ when
both $a \to 0$ and $a \to +\infty$. In this case it has a single minimum value. For the commutative model, if we choose the parameters, such that the minimum of the potential is positive, then the dynamical equations for the scale factor have no real solutions. For the NC model, with $\alpha < 0$ and depending on the value of $\alpha$, the minimum of the potential becomes negative. In those cases, the scale factor will increase up to a maximum value and then it will contract to a positive minimum value. After that, it will continue to oscillate between these maxima and minima values. Therefore, we can see that the presence of a NC parameter may produce a huge change in the evolution of the Universe. An example of this situation is shown in Figure 4.

\begin{figure}
\includegraphics[width=8cm,height=6cm, angle=0]{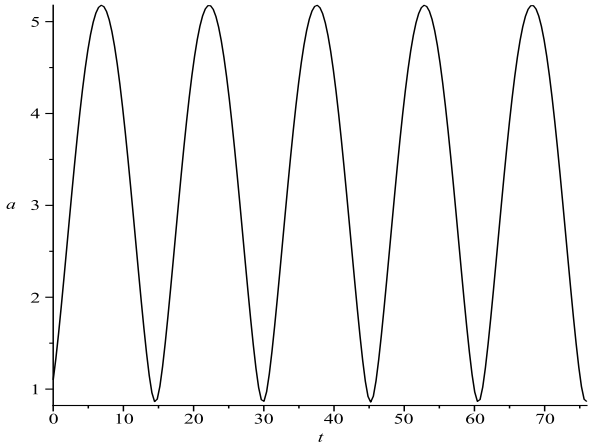}
\caption{Scale factor dynamics of a NC model where the matter is described by cosmic strings and $k = - 1$. The parameters have the following values: $\alpha = -0.5$, $g_{\Lambda} = -0.1$, $g_r = -0.2$, $g_s = 0.1$ and $\Omega = -0.1$. The corresponding commutative case has no real solution for the scale factor.}
\label{fig4}
\end{figure}

\subsubsection{Commutative models without solutions versus NC models with expanding Universes}

In the present case, let us consider another commutative FRW HL cosmological model where the dynamical equations for the scale factor have no real solutions. After the introduction of the NC algebra, as we will see, the scale factor will increase as a function of time. This case is very important because it provides another example of a model where NCY can produce an accelerated expansion of the Universe. For a specific example of this situation, let us consider a model where the geometry is coupled to a cosmological constant ($\omega=-1$) and the constant curvature of the spatial sections is positive ($k=1$). The potential $V(a)$ in Eq. (\ref{22}) is given by,
\begin{equation}
\label{36}
V_{cc}(a) = g_{c} - \Omega a^{2} - \frac{g_{r}}{a^{2}} - \frac{g_{s}}{a^{4}} + \alpha a^{3},
\end{equation}
where $\Omega = g_{\Lambda} + C$ and the potential for the commutative case is obtained by choosing $\alpha=0$ in $V_{cc}$ above. In this model, both commutative and NC ($\alpha < 0$) cases have very different asymptotic behaviors, when $a \to \infty$. Let us choose the parameters such that, $g_s$ and $\Omega$ are negative and $g_{r}$ is positive. For the commutative case, the potential $V_{cc}$ in Eq. (\ref{36}) diverges to $+\infty$ when both $a \to 0$ and $a \to +\infty$. In this case, it has only one  minimum value. If we choose the values of the parameters, such that the minimum of the potential is positive, the dynamical equations for the scale factor have no real solutions. For the NC space-time model, with $\alpha < 0$, the potential $V_{cc}$ in Eq. (\ref{36}) diverges to $+\infty$ when $a \to 0$ but now it diverges to $-\infty$ when $a \to +\infty$. In this case, the potential has an inflection point or a local minimum value. Let us choose the same values of the parameters, as in the commutative case, and an appropriate value of $\alpha$, such that the local minimum of the potential is positive. Therefore, under these conditions the scale factor has just one behavior as a function of time. It will expand in an accelerated rate, from the  initial value, toward a {\it big rip} singularity, in a finite time. Therefore, we see that the NC parameter can introduce a huge modification in the description of 
the evolution of the Universe. In this case, we have a description of the accelerated expansion of the Universe. An example of this situation is shown in Figure 5.

\begin{figure}
\includegraphics[width=8cm,height=6cm, angle=0]{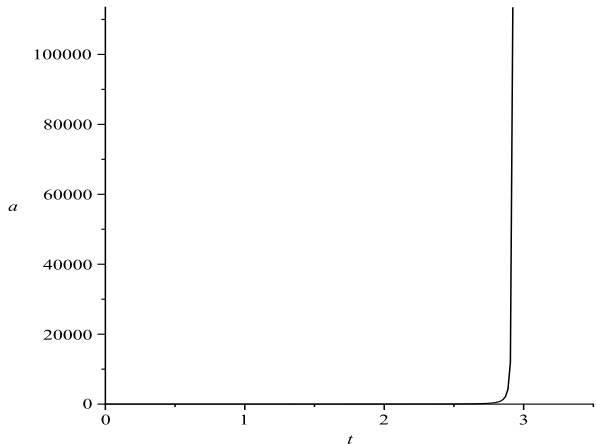}
\caption{Scale factor dynamics of a NC model where the matter is described by cosmological constant and $k = 1$. The parameters have the following values: $\alpha = -0.5$, $g_c = 1.2$, $g_r = 1$, $g_s = - 0.2$, and $\Omega = - 0.6$. The corresponding commutative case has no real solution for the scale factor.}
\label{fig5}
\end{figure}

\section{Estimating the value of $\alpha$}
\label{section5}

We can estimate the value of $\alpha$ by using the equation that gives the age of the Universe \cite{weinberg},
\begin{equation}
\label{37}
t_0 = \int\limits_{0}^{a_{0}}\frac{1}{xH(x)}dx,
\end{equation}
where $t_0$ and $a_0$ are, respectively, the current age and scale factor of the Universe.
With the aid of the generalized Friedmann equation Eq. (\ref{20}), the Hubble parameter $H(a)$ is given by,
\begin{equation}
\label{38}
\frac{\dot{a}^2}{a^2}\equiv H^{2}(a) = - \frac{kg_{c}}{a^2} + g_{\Lambda} + k^{2}\frac{g_{r}}{a^4} + k^{3}\frac{g_{s}}{a^6} + \frac{C}{a^{3\omega + 3}} - \frac{\alpha}{a^{3\omega + 2}}.
\end{equation}
Now, using the value of $H(a)$ given in Eq. (\ref{38}), the equation for the age of the
Universe Eq. (\ref{37}) becomes,
\begin{equation}
\label{39}
t_0 = \int\limits_{0}^{a_0}\frac{1}{x\sqrt{- \frac{kg_{c}}{x^2} + g_{\Lambda} + k^{2}\frac{g_{r}}{x^4} + k^{3}\frac{g_{s}}{x^6} + \frac{C}{x^{3\omega + 3}} - \frac{\alpha}{x^{3\omega + 2}}}}\,dx.
\end{equation}
Now, we will make some simplifying assumptions. Firstly, we consider that the current constant curvatures of
the spatial sections of our Universe are zero, it means that $k = 0$ in Eq. (\ref{39}). Secondly, we
suppose that there is not a cosmological constant at current times.   It means that $g_\Lambda = 0$ in
Eq. (\ref{39}). After these simplifications, Eq. (\ref{39}) can be written in the following way,
\begin{equation}
\label{40}
t_0 = \int\limits_{0}^{a_0}\frac{1}{\sqrt{\frac{C}{x^{3\omega + 1}} - \frac{\alpha}{x^{3\omega}}}}\,dx.
\end{equation}
In order to compute $\alpha$ in Eq. (\ref{40}), we must rewrite $C$ in terms of the observable
quantities. It can be done by setting $C = 8\pi G \rho_0/3$, where $G$ is the gravitational constant and $\rho_0$ is the current matter density of the Universe. Now, if we define the
critical density $\rho_c$ as being $\rho_{c}=3H^{2}/(8\pi G)$, and the current matter density parameter $\Omega_0$ such as $\Omega_0=\rho_{0}/\rho_c$, then Eq. (\ref{40})
will be given by,
\begin{equation}
\label{41}
t_0 = \int\limits_{0}^{a_{0}}\frac{1}  {\sqrt{\Omega_{0} H_{0}^{2}x^{-3\omega-1}
 - \frac{\psi}{3} x^{-3\omega}}}\,dx,
\end{equation}
where we have introduced a new NC parameter $\psi = 3\alpha$.
Finally, let us consider that at current times, the Universe is dominated by a pressureless perfect fluid of dust, which
means that $\omega = 0$. Now, by introducing the following values for the parameters in Eq. (\ref{41}): $\Omega_0 = 0,3$, 
$H_{0} = 72 Km s^{-1} Mpc^{-1}$, $t_0 = 4,32 \times 10^{17} s$ ($\approx 13,7 \times 10^9$ years) and $a_0 = 1$, this
equation may be written as,
\begin{equation}
\label{42}
\int\limits_{0}^{1}\frac{1}{\sqrt{(1,538)10^{-36}x^{-1}-\frac{\psi}{3}}}dx = 4,32 \,\times\, 10^{17}.
\end{equation}

That equation (\ref{42}) is identical to  Eq. (6.6), obtained in Ref. \cite{mateus}.
In this paper, some of us solved, numerically, Eq. (\ref{42}) and found that $\psi \approx -0,45 \times 10^{-35} s^{-2}$. 
Following the comparison, made in Ref. \cite{mateus}, between the energy density of the real perfect fluid and the NC perfect fluid, it is important to notice that for large values of $a$ and any value of $\omega$, the ``energy density" of the noncommutative fluid $\vartheta$ Eq. (\ref{28}), dominates. For small values of $a$ and any value of $\omega$, the energy density of the real fluid $\rho$ Eq. (\ref{26}), dominates. It is also observed, in Ref. \cite{mateus}, that the numerical value 
of $\alpha$ is of the same order of magnitude of the presently estimated value of the cosmological constant $g_\Lambda$, in the HL theory. From the literature, $g_{\Lambda 0} \approx 2.28 \times 10^{-35} s^{-2}$ \cite{carmeli}.
\section{Final considerations and conclusions}
\label{section6}




In this paper, we have explored a NC version of FRW cosmological models in the gravitational HL theory. The matter content of the models is described by a perfect fluid and the constant curvature of the spatial sections may be positive, negative or zero. 

To obtain this theory, we have used the FJ formalism to introduce, naturally, a NC algebra inside the equations that describe the dynamics of the theory. Using this formalism we have obtained the corresponding NC Lagrangian and Hamiltonian densities of the theory. In the most general formulation, obtained here via the FJ formalism, the theory has three different NC parameters. We have recovered the commutative FRW HL cosmological models by setting those NC parameters to zero. Due to the difficulty to work with the most general NC theory, we have concentrated our attention, in the explicit calculations of the model dynamics, to the case with just one NC parameter. This kind of NCY has already been analyzed in the literature, in the context of general relativistic FRW cosmological models \cite{mateus}. We have derived the dynamical equations for the scalar factor for different models. With these equations, we have showed that the energy
conservation still holds, for those NC models. We have also proved that the NC term, present in the dynamical equations, may be understood as being originated from a perfect fluid with a specific equation of state.


After solving the dynamical equations for many different models, we concluded that, if the parameter $\alpha$ is positive, it has the net effect of making the Universe expansion more difficult. In cases where the Universe is expanding, the presence of a positive $\alpha$, will slow the expansion down or even stops it and forces the Universe to contract. If the parameter $\alpha$ is negative, it has the net effect of making the Universe
expansion easier. In cases where the Universe is expanding, the presence of a
negative $\alpha$, will increase the expansion speed. On the other hand, in cases where the Universe is contracting, the presence of a negative $\alpha$ may force the Universe to expand. We have also presented some particular cases where the above conclusions could be clearly verified. 

Finally, we estimated the value of the noncommutative parameter $\alpha$, with the aid of a result already published in the literature \cite{mateus}.
Since we are particularly interested in describing the present expansion of our Universe, we may mention that, due to 
the noncommutativity introduced here, we have an extra free parameter $\alpha$ (specially in the case of a negative $\alpha$) not present in the corresponding
commutative models. One may use that extra freedom to better adjust the observational data. 
\acknowledgments 

L.G. Rezende Rodrigues and M. Silva de Oliveira thank CAPES for their scholarships.   E.M.C.A.  and J.A.N. thank CNPq (Conselho Nacional de Desenvolvimento Cient\' ifico e Tecnol\'ogico), Brazilian scientific support federal agency, for partial financial support, Grants numbers 302155/2015-5 (E.M.C.A.) and 303140/2017-8 (J.A.N.). E.M.C.A. thanks the hospitality of Theoretical Physics Department at Federal University of Rio de Janeiro (UFRJ), where part of this work was carried out.


\end{document}